\documentclass[%
pre,preprint,superscriptaddress,
tightenlines,
showpacs,
a4paper,12pt
]{revtex4}
\usepackage{amssymb,amsmath}
\usepackage{graphicx}

\newcommand{\sca}[2]{\ensuremath{\bigl({#1},{#2}\bigr)}}

\newcommand{\hcnj}[1]{{#1}^{+}}

\newcommand{\bs}[1]{\boldsymbol{#1}}
\newcommand{\vc}[1]{\mathbf{#1}}
\newcommand{\uvc}[1]{\mathbf{\hat #1}}

\newcommand{\dd}{\mathrm{d}}

\newcommand{\prt}[1]{\partial_{#1}}
\newcommand{\ddiv}{\mathop{\rm div}\nolimits}
\newcommand{\bnbl}{\boldsymbol{\nabla}}

\begin{document}

 \title{Twist of
cholesteric liquid crystal cells:\\
stability of helical structures and anchoring energy effects
}

\author{A.D.~Kiselev}
\email[Email address: ]{kisel@mail.cn.ua}
\affiliation{%
 Chernigov State Technological University,
 Shevchenko Street 95,
 14027 Chernigov, Ukraine
}
\author{T.J.~Sluckin}
\email[Email address: ]{t.j.sluckin@maths.soton.ac.uk}
\affiliation{%
 School of Mathematics,
 University of Southampton,
 Southampton, SO17 1BJ, United Kingdom
 }

\date{\today}

\begin{abstract}
We consider helical configurations of a cholesteric liquid crystal
(CLC) sandwiched between two substrates with homogeneous director
orientation favored at both confining plates.  We study the CLC
twist wavenumber $q$ characterizing the helical structures in
relation to the free twisting number $q_0$ which determines the
equilibrium value of CLC pitch, $P_0=2\pi/q_0$. We investigate the
instability mechanism underlying  transitions between helical
structures with different spiral half-turn numbers. Stability
analysis  shows that for equal finite anchoring strengths this
mechanism can be dominated by  in-plane director fluctuations.  In
this case the metastable helical configurations are separated by
the energy barriers and the transitions can be described as the
director slippage through these barriers.  We extend our analysis
to the case of an asymmetric CLC cell in which the anchoring
strengths at the two substrates are different. The asymmetry
introduces two qualitatively novel effects: (a)~the intervals of
twist wavenumbers representing locally stable configurations with
adjacent helix half-turn numbers are now separated by the
instability gaps; and (b)~sufficiently large asymmetry, when the
difference between azimuthal anchoring extrapolation lengths
exceeds the thickness of the cell, will suppress the jump-like
behaviour of the twist wavenumber.
\end{abstract}



\maketitle

\section{Introduction}
\label{sec:intro}

In equilibrium cholesteric phase molecules of a liquid crystal (LC)
align on average along a local unit director $\vc{n}(\vc{r})$ that
rotates in a helical fashion about a uniform twist
axis~\cite{Gennes:bk:1993}.  This tendency of cholesteric
liquid crystals (CLC) to form helical twisting patterns
is caused by the presence of anisotropic molecules with
no mirror plane~---~so-called chiral molecules
(see~\cite{Harr:rmp:1999} for a recent review).

The phenomenology of CLCs can be explained in terms of
the Frank free energy density
\begin{align}
& f_b[\vc{n}]=\frac{1}{2} \Bigl\{ K_1 ({\nabla}\cdot\vc{n})^2+K_2
\left[ \vc{n}\cdot{\nabla\times\vc{n}}+q_0 \right]^2 \notag\\ &
+K_3\,[\vc{n}\times(\nabla\times\vc{n})]^2 -K_{24} \ddiv\left(
\vc{n}\ddiv\vc{n}+ \vc{n}\times(\nabla\times\vc{n}) \right)
\Bigr\}\,, \label{eq:frank}
\end{align}
where $K_1$, $K_2$, $K_3$ and $K_{24} $ are the splay, twist, bend
and saddle-splay Frank elastic constants.
As an immediate consequence of the broken mirror symmetry,
the expression for the bulk free energy~\eqref{eq:frank}
contains a chiral term proportional
to the equilibrium value of the CLC twist wavenumber, $q_0$.

The parameter $q_0$, which will be referred to as
the free twist wavenumber or as the free twisting number,
gives the equilibrium helical pitch
$P_0\equiv 2\pi/q_0$.
For the twist axis directed along the
$z$-direction, the director field
$\uvc{n}=(\cos q_0 z, \sin q_0 z, 0)$ then defines the equilibrium
configuration in an unbounded CLC.
Periodicity of the spiral is given by the half-pitch, $P_0/2$,
because $\uvc{n}$ and $-\uvc{n}$ are equivalent in liquid crystals,

Typically, the pitch $P_0$ can vary from hundreds of nanometers to
many microns or more, depending on the system.  The macroscopic
chiral parameter, $h=q_0 K_2$, (and thus the pitch) is determined
by microscopic intermolecular torques~\cite{Harr:1997,Luben:1997}
and depends on the molecular chirality of CLC consistuent
mesogens. The microscopic calculations of the chiral parameter are
complicated as it is necessary to go beyond the mean-field
approach and to take into account biaxial
correlations~\cite{Harr:rmp:1999}. Despite  recent
progress~\cite{Osip:pre:2000,Emel:pre:2003}, this problem has not
been resolved completely yet.

In this paper we are primarily concerned with orientational structures
in planar CLC cells bounded by two parallel substrates.
Director configurations in such cells are strongly affected by the
anchoring conditions at the boundary surfaces
which break the translational symmetry along the twisting axis.
So, in general, the helical form of the director field will be distorted.

Nevertheless, when the anchoring conditions are planar and
out-of-plane deviations of the director are suppressed, it might
be expected that the configurations still have the form of the
ideal helical structure. But, by contrast with  the case of
unbounded CLCs, the helix twist wavenumber $q$ will now differ
from $q_0$.

It has long been known that a mismatch between the equilibrium pitch
$P_0$ and the twist imposed by the boundary conditions may produce two
metastable twisting states that are degenerate in energy and can be
switched either way by applying an electric
field~\cite{Berrem:jap:1981}.  This bistability underlines the mode of
operation of bistable liquid crystal devices~---~the so-called
bistable twisted nematics~---~that have been attracted considerable
attention over past few
decades~\cite{Xie:jap:1998,Zhuang:apl:1999,Xie:jap2:2000,Kwok:apl:2003}.

More generally the metastable twisting states in CLC cells appear
as a result of interplay between the bulk and the surface
contributions to the free energy giving rise to multiple local minima
of the energy.  The purpose of this paper is to explore the multiple minima
and their consequences.

The free twisting number $q_0$ and the anchoring energy are among
the factors that govern the properties of the multiple minima
representing metastable states. Specifically, varying $q_0$ will
change the twist wavenumber of the twisting state, $q$. This may
result in sharp transitions between different branches of
metastable states.  The dependence of the twist wavenumber $q$ on
the free twisting number $q_0$ is then discontinuous. As far as we
are aware, attention was first drawn to this phenomenon by
Reshetnyak~\textit{et. al}~\cite{Pink:1992}.

These discontinuities are accompanied by a variety of physical
manifestations which have been the subject of much recent important
research. One such is a jump-like functional dependence of selective
light transmission spectra on temperature as a result of a
temperature-dependent cholesteric pitch, examined by Zink and
Belyakov~\cite{Zink:jetp:1997,Zink:1999}.
More recently Belyakov~\textit{et. al}~\cite{Bel:eng:2000,Bel:eng:2003}
and Palto~\cite{Palto:eng:2002} have discussed
different mechanisms behind temperature variations of the pitch in CLC cells and hysteresis
phenomena.

In this paper we adapt a systematic approach and
study the helical structures using stability analysis.
This approach enables us to go beyond the previous work by relaxing a
number of constraints. One of these requires anchoring to be
sufficiently weak (where ``sufficiently'' will be discussed further below),
so that the jumps may occur only due to
transitions between the helical configurations which numbers
of spiral half-turns differ by the
unity~\cite{Bel:eng:2000,Bel:eng:2003}.
Noticeably, this assumption eliminates important class of
the transitions  that involve topologically equivalent structures with the half-turn
numbers of the same parity.

We shall also apply our theory to the case of non-identical confining
plate and show that asymmetry in the anchoring properties of the
bounding surfaces results in qualitatively new effects.
Specifically, we find that sufficiently large asymmetry in anchoring
strengths will suppress the jump-like behaviour of the twist
wavenumber $q$ when the free wavenumber $q_0$ varies.

The layout of the paper is as follows.  General relations that
determine the characteristics of the helical structures in CLC cells
are given in Sec.~\ref{sec:general}. Then in Sec.~\ref{sec:stability}
we outline the procedures which we use to study stability of the director
configurations. The stability analysis is performed for in-plane and out-of-plane
fluctuations invariant with respect to in-plane
translations.  We study CLC cells with the strong anchoring conditions
and the cases where at least one anchoring strength is finite.
We formulate the stability conditions and the criterion for the stability of the helical
structures to be solely governed by the in-plane director
fluctuations. The expressions for the fluctuation static
correlation functions are given.
In Sec.~\ref{sec:pitch-vs-twist}
we study the dependence of the twist wavenumber on the free twisting
number.
Finally, in Sec.~\ref{sec:concl} we present our results and
make some concluding remarks.
Details on some technical results are relegated to
Appendix~\ref{sec:stab-fluct-spectr}.

\section{Helical structures}
\label{sec:general}

\subsection{Energy}
\label{subsec:energy}

We consider a CLC cell of thickness $d$ sandwiched
between two parallel plates that are normal to
the $z$-axis:  $z=-d/2$ and $z=d/2$.
Anchoring conditions at both substrates are planar
with the preferred orientation of CLC molecules at the lower and upper
plates defined by
the two vectors of easy orientation: $\uvc{e}_{-}$
and $\uvc{e}_{+}$. These vectors are given by
\begin{equation}
  \label{eq:clc-anch-vec}
  \vc{e}_{-}=\vc{e}_x,\quad
\vc{e}_{+}=\cos\Delta\phi\,\vc{e}_x+\sin\Delta\phi\,\vc{e}_y,
\end{equation}
where $\Delta\phi$ is the twist angle imposed by the boundary conditions.

We shall also write the elastic free energy as a sum of the bulk and
surface contributions:
\begin{align}
 F[\vc{n}]=\int _V f_b[\vc{n}]\,\dd v +\sum_{\nu=\pm 1}
\int_{z=\nu d/2} W_{\nu}(\vc{n})\,\dd s
\label{eq:f-gen}
\end{align}
and assume that both the polar and the azimuthal
contributions to the anchoring energy $W_{\nu}(\vc{n})$
can be taken in the form of Rapini-Papoular potential~\cite{Rap:1969}:
\begin{align}
W_{\nu}(\vc{n})=
\frac{W^{(\nu)}_{\phi}}{2}
\Bigl[1-\sca{\vc{n}}{\vc{e}_{\nu}}^2\Bigr]_{z=\nu d/2}
+
\frac{W^{(\nu)}_{\theta}}{2}
\Bigl[1-\sca{\vc{n}}{\vc{e}_{z}}^2\Bigr]_{z=\nu d/2},
  \label{eq:f_s-gen}
\end{align}
where $W^{(\pm)}_{\phi}$ and $W^{(\pm)}_{\theta}$ are the azimuthal
and the polar anchoring strengths.

The CLC helical director structures take the following spiral form
\begin{equation}
  \label{eq:n0}
  \vc{n}_0=\cos u(z)\,\vc{e}_x+\sin u(z)\,\vc{e}_y,\quad
u(z)=q z + \phi_0,
\end{equation}
where $q$ is the twist (or pitch) wavenumber and
$\phi_0$ is the twist angle of the director in the middle of the cell.
The configurations~\eqref{eq:n0} can be obtained as a solution
of the Euler-Lagrange equations for the free energy
functional~\eqref{eq:frank} provided the invariance with respect
to translations in the $x-y$ plane is unbroken.

The translation invariant solutions can be complicated by the presence of the
out-of-plane director deviations neglected in Eq.~\eqref{eq:n0} and,
in general, does not represent a helical structure.  Using
Eq.~\eqref{eq:n0} is justified only for those configurations that
are stable with respect to out-of-plane director fluctuations.  The
corresponding stability conditions will be derived in the next
section.

We can now substitute Eq.~\eqref{eq:n0} into
Eq.~\eqref{eq:f-gen} to obtain
the following expression for the rescaled free energy per unit area
of the director configuration~\eqref{eq:n0}:
\begin{equation}
  \label{eq:en-n0}
  (2d/K_2) F[\vc{n}_0]\equiv f[\vc{n}_0] =
(q d - q_0 d)^2+ 2 \sum_{\nu=\pm 1} w^{(\nu)}_{\phi}\sin^2 u_{\nu},
\end{equation}
where $u_{\nu}$ is the angle between the vector of easy orientation
$\vc{e}_{\nu}$ and the director $\vc{n}_0$ at the plate
$z=\nu d/2$; and
the dimensionless azimuthal anchoring energy parameter
$w^{(\nu)}_{\phi}$ is proportional to the ratio of the cell thickness, $d$,
and the azimuthal extrapolation length, $L^{(\nu)}_{\phi}=W^{(\nu)}_{\phi}/K_2$:
\begin{equation}
  \label{eq:par-w}
  w^{(\nu)}_{\phi}\equiv \frac{W^{(\nu)}_{\phi} d}{2K_2}
=\frac{d}{2 L^{(\nu)}_{\phi}}, \quad
\cos u_{\nu}\equiv \sca{\vc{n}_0}{\vc{e}_{\nu}}\Bigr\vert_{z=\nu d/2}.
\end{equation}

The energy~\eqref{eq:en-n0} is of the well-known ``smectic-like''
form~\cite{Gennes:bk:1993} and can be conveniently rewritten
in terms of the following dimensionless parameters
\begin{equation}
  \label{eq:par-bet}
  \beta=q d - \Delta\phi,\quad
  \beta_0=q_0 d - \Delta\phi,\quad
  \alpha=2\phi_0 - \Delta\phi
\end{equation}
by using the relations
\begin{equation}
  \label{eq:par-unu}
d(q-q_0)=\beta-\beta_0,\quad
  2 u_{\nu} = \nu\beta+\alpha.
\end{equation}
Given the free twist parameter $\beta_0$
the energy~\eqref{eq:en-n0} is now a function of
$\alpha$ and the twist parameter $\beta$ which
characterize the helical structure~\eqref{eq:n0}.
The azimuthal angles of the director at the bounding surfaces,
$\phi_{\nu}=u(\nu d/2)$, can be expressed in terms of the parameters~\eqref{eq:par-bet}
and $\Delta\phi$ as follows
\begin{equation}
  \label{eq:phi-nu}
  \phi_{+}=u_{+}+\Delta\phi=(\beta+\alpha)/2+\Delta\phi,\quad
\phi_{-}=u_{-}=(-\beta+\alpha)/2.
\end{equation}

\subsection{Twist wavenumber and parity}
\label{subsec:wavenumber}

It is not difficult to show that
in order for the configuration to be an extremal of the free
energy~\eqref{eq:f-gen} these parameters need to satisfy
the system of the following two equations:
\begin{equation}
  \label{eq:extr}
  \beta_0=\beta + w^{(\nu)}_{\phi}\sin(\beta+\nu\alpha),\quad
\nu=\pm 1.
\end{equation}
Equivalently, this system determines the extremals as
stationary points of the energy~\eqref{eq:en-n0}
and can be derived from the condition that
both energy derivatives with respect to $\alpha$ and $\beta$
vanish.

Eq.~\eqref{eq:extr} can now be used to relate the parameters
$\alpha$ and $\beta$ through the equation
\begin{equation}
  \label{eq:alp}
  \alpha=\arctan[\epsilon \tan\beta]+\pi k,\quad
\epsilon\equiv\frac{w^{(-)}_{\phi}-w^{(+)}_{\phi}}{w^{(-)}_{\phi}+w^{(+)}_{\phi}},
\end{equation}
where $k$ is the integer, $k\in \mathbb{Z}$,
that defines the parity of the configuration $\mu = (-1)^k$.

Indeed, substituting Eq.~\eqref{eq:alp} into
Eq.~\eqref{eq:extr} gives the relation between
$\beta_0$ and $\beta$
\begin{equation}
  \label{eq:extr2}
  \beta_0=\gamma_{\mu}(\beta)\equiv
\beta+\mu w^{(+)}_{\phi} \sin(\beta+\arctan[\epsilon \tan\beta]),\quad
\mu = (-1)^k,
\end{equation}
that depends on $k$ only through the parity.
This remark also applies to the expression for the
energy that after substituting the
relation~\eqref{eq:alp} into Eq.~\eqref{eq:en-n0}
can be recast into the form
\begin{equation}
  \label{eq:fmu}
  f_{\mu}(\beta)=\bigl[w^{(+)}_{\phi}\sin v_{+}\bigr]^2+(w^{(+)}_{\phi}+w^{(-)}_{\phi})-\mu
\sum_{\nu=\pm 1}w^{(\nu)}_{\phi}\cos v_{\nu},
\end{equation}
where
\begin{equation}
  \label{eq:par-vnu}
  v_{\nu}=\beta+\nu\arctan[\epsilon \tan\beta].
\end{equation}
In Sec.~\ref{sec:pitch-vs-twist} we will find that there are different
branches of metastable helical configurations.
Each branch is characterized by the
number of the spiral half-turns and $\mu$ is the parity
of this number. For this reason, the integer $k$ will be referred to as the
half-turn number.

\begin{figure*}[!tbh]
\centering
\resizebox{150mm}{!}{\includegraphics*{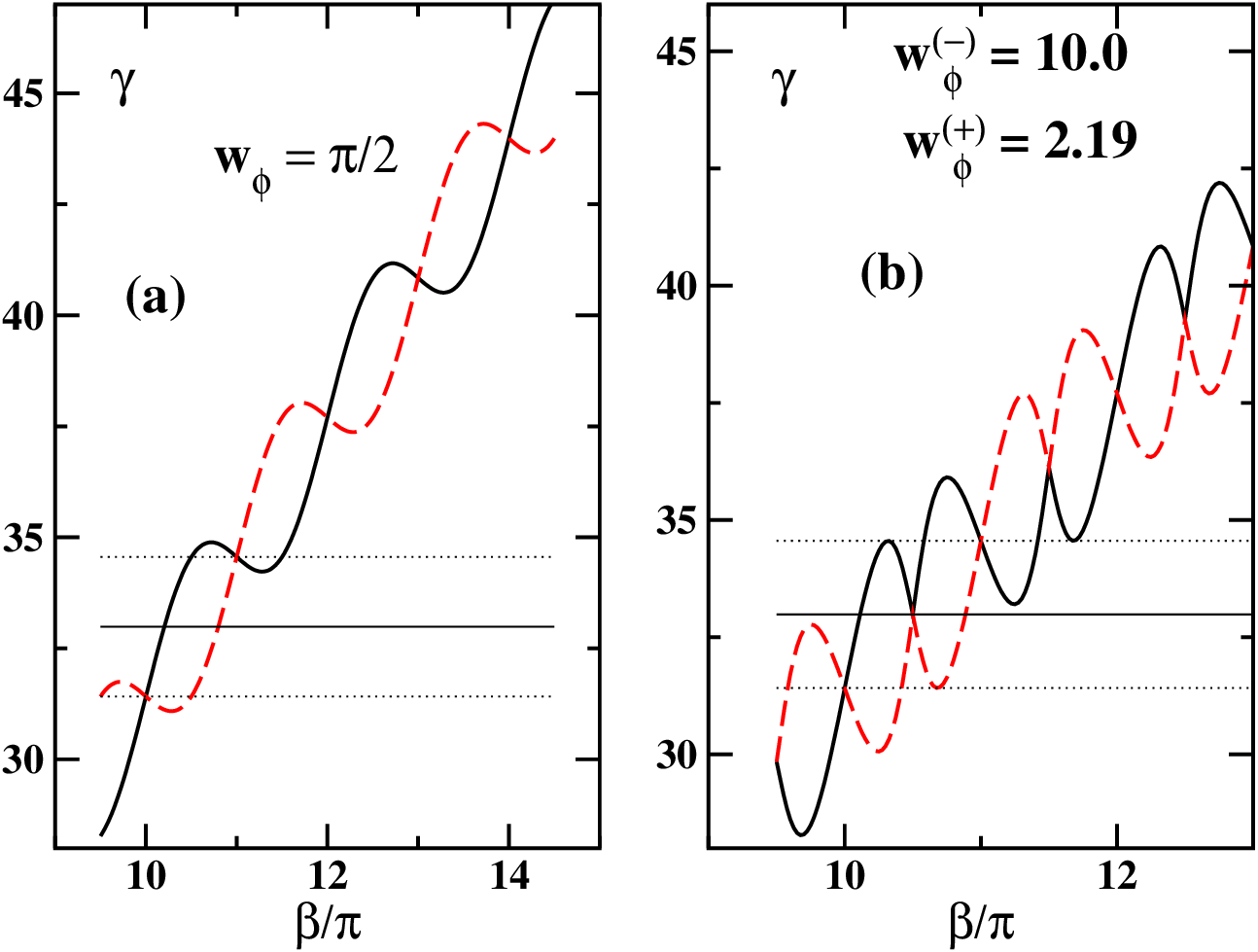}}
\caption{%
The curves
representing the plot of the function
$\gamma_{+}(\beta)$ and $\gamma_{-}(\beta)$
are shown as thick solid and dashed lines,
respectively.
The points located at the intersection of
the curves and the horizontal straight line
$\gamma=\beta_0$ give the roots of Eq.~\eqref{eq:extr2}.
The value of $\beta_0$ is $(10+1/2)\pi$
(thin solid line) and is $(10+1/2)\pi\pm w_{\phi}$
(thin dotted lines).
Two cases are illustrated:
(a) $w^{(+)}_{\phi}=w^{(-)}_{\phi}=w_{\phi}=\pi/2$;
(b) $w^{(-)}_{\phi}=10.0$, $w^{(+)}_{\phi}=2.19$,
$w_{\phi}=\pi/2$ (see Eq.~\eqref{eq:ww}).
}
\label{fig:inter}
\end{figure*}

Thus, we have classified the director structures by means of the
parity $\mu$ and the dimensionless twist parameter $\beta$ that can be
computed by solving the transcendental equation~\eqref{eq:extr2}.
Fig.~\ref{fig:inter} illustrates the procedure of finding the roots of
Eq.~\eqref{eq:extr2} in the $\beta-\gamma$ plane.

In general, there are several roots represented by the intersection
points of the horizontal line $\gamma=\beta_0$ and the curves
$\gamma=\gamma_{\pm}(\beta)$.
Each root corresponds to the director
configuration which energy can be calculated from Eq.~\eqref{eq:fmu}.
The equilibrium director structure is then determined by the solution of
Eq.~\eqref{eq:extr2} with the lowest energy.
Other structures can be either metastable or unstable.

\subsection{Strong anchoring limit}
\label{subsubsec:strong-anch-limit}

However, these results cannot be applied directly to the case of the strong anchoring
limit, where
$W^{(\nu)}_{\phi}\to\infty$ and the boundary condition requires the
director at the substrate $z=\nu d/2$ to be parallel to the corresponding easy
axis, $\vc{n}(\nu d/2)\parallel \vc{e}_{\nu}$.

When the anchoring is strong at both substrates,
it imposes the
restriction on the values of $q$, so that $q$ takes the values from
a discrete set~\cite{Gennes:bk:1993}. This set represents the director
configurations characterized by the parameter $\beta$ and
labelled by the half-turn number $k$
\begin{equation}
\beta\equiv qd-\Delta\phi=\pi k,\quad k\in\mathbb{Z}\,.
  \label{eq:qd-str}
\end{equation}

\begin{figure*}[!tbh]
\centering
\resizebox{150mm}{!}{\includegraphics*{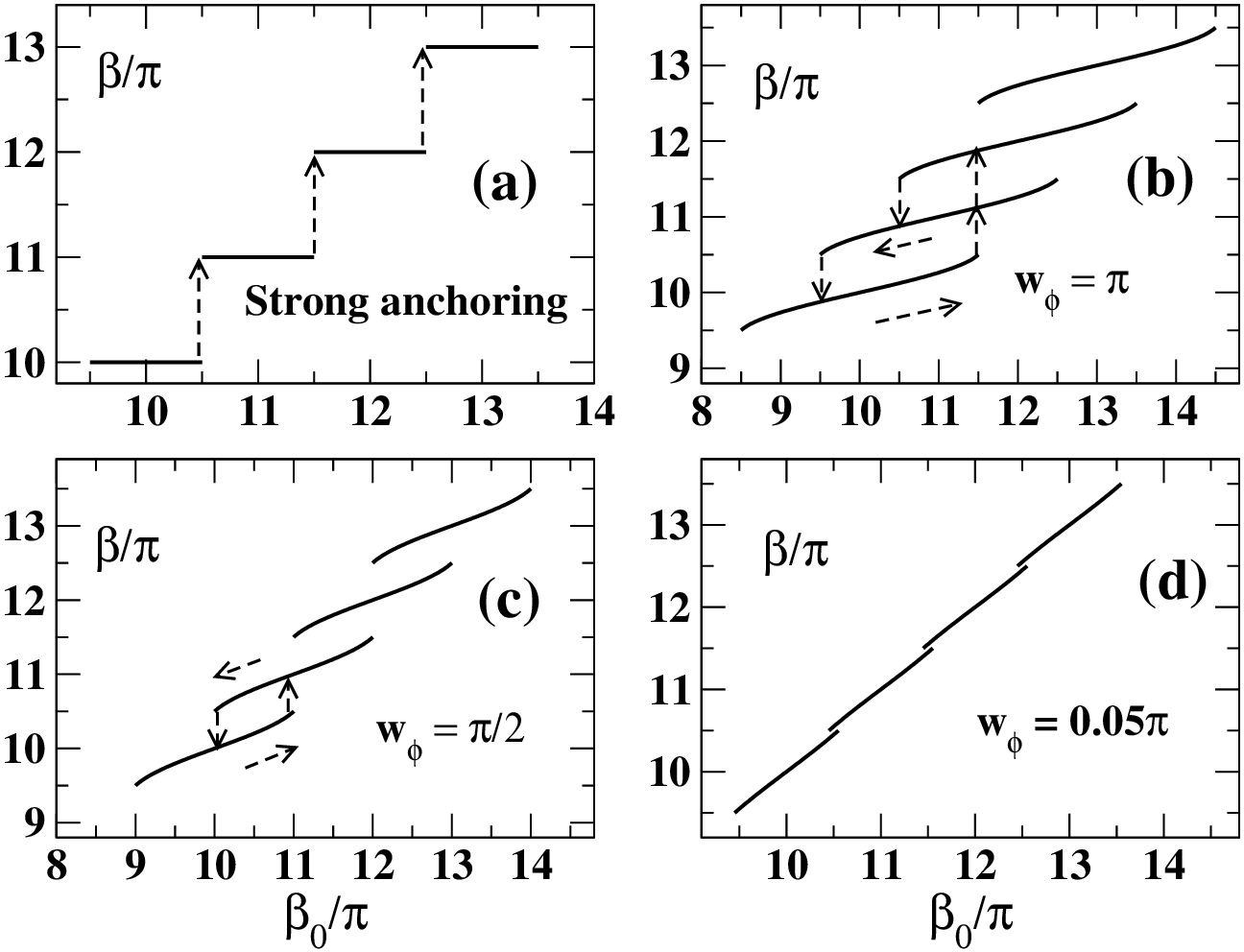}}
\caption{%
Dependence of the twist parameter $\beta$ ($=qd-\Delta\phi$) on
the free twisting parameter $\beta_0$
($=q_{0}d-\Delta\phi$) at $W^{(-)}_{\phi}=W^{(+)}_{\phi}\equiv W_{\phi}$
for various values of the dimensionless azimuthal anchoring parameter
$w_{\phi}$ ($=W_{\phi}d/2K_2$): (a) strong anchoring limit,
$w_{\phi}\to\infty$, discussed in  Sec.~\ref{subsubsec:strong-anch-limit};
(b) $w_{\phi}=\pi$; (c) $w_{\phi}=\pi/2$; (d) $w_{\phi}=0.05\pi$.
Discussion of the cases (b)-(d) can be found in Sec.~\ref{subsec:equal-anch}.
}
\label{fig:sym}
\end{figure*}

Substituting the values of
$\beta$ from Eq.~\eqref{eq:qd-str} into the first term
on the right hand side of Eq.~\eqref{eq:en-n0} will define
the equilibrium value of $k$ as the integer  that minimizes the distance
between $\pi k$ and $\beta_0$ ($=q_0 d - \Delta\phi$).
The step-like dependence of $\beta$ on $\beta_0$
for these equilibrium structures is depicted in Fig.~\ref{fig:sym}(a).

An experimentally important case concerns mixed boundary conditions in which the strong
anchoring limit applies to the lower plate only,
$W^{(-)}_{\phi}\to\infty$. For brevity, this case will be referred to as the semi-strong anchoring.
Now the relations~\eqref{eq:alp}
and~\eqref{eq:extr2} reduce to
\begin{align}
&
  \label{eq:alp-semi}
  \alpha=\beta+2\pi k,
\\
&
  \beta_0=
\beta+w^{(+)}_{\phi} \sin(2 \beta)
\label{eq:beta0-semi}
\end{align}
and the energy of the helical structures~\eqref{eq:fmu} is now given
by
\begin{equation}
  \label{eq:fmu-semi}
f(\beta)=\bigl[w^{(+)}_{\phi}\sin (2\beta)\bigr]^2+
2 w^{(+)}_{\phi}\sin^2 \beta.
\end{equation}
Interestingly, in the semi-strong anchoring limit,
the parity of half-turns, $\mu$, does not enter either the
energy~\eqref{eq:fmu-semi} or the relation~\eqref{eq:beta0-semi}.

\section{Stability of helical structures}
\label{sec:stability}

In this section we present the results on
stability of the helical configurations~\eqref{eq:n0}.
These results then will be used in Sec.~\ref{sec:pitch-vs-twist}
to eliminate unstable structures from consideration.
We also give expressions for director correlation
functions. These are in order
to discuss the effects of director fluctuations.

We begin with the general expression for the distorted director field
\begin{equation}
  \label{eq:dir-gen}
  \vc{n}=\cos\theta\cos\phi\,\vc{n}_0+
\cos\theta\sin\phi\,\vc{n}_1+\sin\theta\,\vc{n}_2,\quad
\sca{\vc{n}_i}{\vc{n}_j}=\delta_{ij},
\end{equation}
where the vectors $\vc{n}_1$ and $\vc{n}_2$ are
\begin{equation}
  \label{eq:n12}
\vc{n}_1=-\sin u(z)\,\vc{e}_x+\cos u(z)\,\vc{e}_y,\quad
\vc{n}_2=\vc{e}_z.
\end{equation}
For small angles $\phi$ and $\theta$
linearization of Eq.~\eqref{eq:dir-gen} gives the perturbed director field
in the familiar form
\begin{equation}
  \label{eq:dir-pert}
  \vc{n}\approx\vc{n}_0+\delta\vc{n}_0,\quad
\delta\vc{n}_0=\phi\,\vc{n}_1+\theta\,\vc{n}_2,
\quad
\bs{\psi}\equiv
\begin{pmatrix}\phi\\ \theta \end{pmatrix},
\end{equation}
where the angles $\phi$ and $\theta$ describe in-plane and out-of-plane
deviations of the director, respectively.

Following standard procedure,
we can now expand the free energy of
the director field~\eqref{eq:dir-gen} up to second order
terms in the fluctuation field $\bs{\psi}$ and its derivatives
\begin{align}
  \label{eq:harm-1}
&
  F[\vc{n}]\approx F[\vc{n}_0]+F^{(2)}[\bs{\psi}],
\\
  \label{eq:harm-2}
&
  F^{(2)}[\bs{\psi}]=\int_V f_{b}^{(2)}[\bs{\psi}]\,\dd v+
\sum_{\nu=\pm 1}
\int_{z=\nu d/2}W_{\nu}^{(2)}(\bs{\psi})\, \dd s.
\end{align}
The second order variation of the free energy $F^{(2)}[\bs{\psi}]$ is
a bilinear functional which represents the energy of the
director fluctuations written in the harmonic (Gaussian) approximation.
>From Eqs.~\eqref{eq:frank}-\eqref{eq:f-gen} we obtain
expressions for the densities that enter
the fluctuation energy~\eqref{eq:harm-2}:
\begin{align}
\label{eq:F2b}
&
2 f^{(2)}_b[\bs{\psi}]=
K_1\bigl(\nabla_1\phi+\nabla_2\theta
\bigr)^2+
K_2 \bigl(\nabla_1\theta-\nabla_2\phi
\bigr)^2
\notag\\
&
+ K_3\bigl[(\nabla_0\phi)^2+(\nabla_0\theta)^2
\bigr]+ q K_q
\bigl[ 2 \theta\,\nabla_0\phi+q \theta^2
\bigr],
\\
\label{eq:F2s}
&
2 W^{(2)}_{\nu}(\bs{\psi})=
\Bigl[
W^{(\nu)}_{\phi}\phi^2\cos 2 u_{\nu}+W^{(\nu)}_{\theta}\theta^2
\notag\\
&
-2\nu K_{24}\,\theta\,\nabla_1\phi
\Bigr]_{z=\nu d/2},\;
\end{align}
where $\nabla_i\equiv\sca{\vc{n}_i}{\bnbl}$.

In what follows we shall restrict our consideration to the case
of fluctuations invariant with respect to in-plane
translations, so that $\bs{\psi}\equiv\bs{\psi}(z)$.
This assumption, although restricting applicability of our results,
allows us to avoid complications introduced by inhomogeneity
of the helical structure~\eqref{eq:n0}.
In this case the fluctuation energy per unit area is
\begin{align}
2F^{(2)}[\bs{\psi}]/S=
\int_{-d/2}^{d/2}
\hcnj{\bs{\psi}}\, \hat{K}
\bs{\psi}\, \dd z
+\sum_{\nu=\pm 1}
\hcnj{\bs{\psi}}\, \hat{Q}^{(\nu)}
\bs{\psi}\Bigr\vert_{z=\nu d/2},
 \label{eq:F2z}
\end{align}
where $S$ is the area of the substrates. The operator
$\hat{K}$ is the differential matrix operator that
enters the linearized Euler-Lagrange equations for the director
distribution~\eqref{eq:dir-gen}, i.e.
\begin{equation}
  \label{eq:lin-EU}
  \hat{K}\bs{\psi}=0.
\end{equation}
The eigenvalues of $\hat{K}$ form the fluctuation spectrum~\cite{Kis:pre:2004}.
The eigenvalues $\lambda$ can be computed together with the eigenmodes
$\bs{\psi}_{\lambda}$  by
solving the boundary-value problem:
\begin{align}
  \label{eq:gen-EV}
  \hat{K}\bs{\psi}_{\lambda}=\lambda\bs{\psi}_{\lambda},
\\
\label{eq:gen-EV-bc}
\hat{Q}^{(\nu)}
\bs{\psi}_{\lambda} \Bigr\vert_{z=\nu d/2}=0.
\end{align}

The expressions for $\hat{K}$ and $\hat{Q}^{(\nu)}$ are given by
\begin{align}
  \label{eq:op-K}
  \hat{K}=
\begin{pmatrix}
-K_2\prt{z}^2 & 0\\
0 & -K_1 \prt{z}^2+q^2 K_q
\end{pmatrix},
\end{align}
\begin{align}
  \label{eq:op-Q}
 \hat{Q}^{(\nu)}=&
\nu
\begin{pmatrix}
K_2\prt{z} & 0\\
0 & K_1 \prt{z}
\end{pmatrix}
\notag\\
&
+\begin{pmatrix}W^{(\nu)}_{\phi}\cos 2 u_{\nu} & 0\\0 & W^{(\nu)}_{\theta}\end{pmatrix},
\end{align}
and $K_q$ is the effective elastic constant
\begin{equation}
  \label{eq:Kq}
  K_q = K_3 - 2 K_2 (1-q_0/q).
\end{equation}
>From Eqs.~\eqref{eq:op-K}-\eqref{eq:op-Q}
the operators $\hat{K}$ and $\hat{Q}^{(\nu)}$ are both diagonal,
so that the in-plane and out-of-plane fluctuations are statistically
independent and can be treated separately.

\subsection{In-plane fluctuations}
\label{subsec:in-plane}

\subsubsection{Strong and semi-strong anchoring}
\label{subsubsec:strong-anch-in}

We begin with the limiting cases discussed in Sec.~\ref{subsubsec:strong-anch-limit}
and where at least one of the bounding surfaces
imposes the strong anchoring boundary condition.
This is the case when we have to use
stability criterion related to the fluctuation spectrum which
requires all the eigenvalues to be positive, $\lambda>0$,
as to ensure the positive definiteness of the fluctuation
energy~\cite{Kis:pre:2004}.

It is not difficult to see that, for the strong azimuthal anchoring present
at both substrates, the lowest eigenvalue is
\begin{equation}
  \label{eq:lowest_eign_value}
  \lambda_m=K_2 (\pi/d)^2
\end{equation}
and all the structures with the twist parameter~\eqref{eq:qd-str}
are locally stable with respect to in-plane fluctuations.

For the semi-strong anchoring with $W^{(-)}_{\phi}\to\infty$, this is no longer the case.
The stability condition is now given by
\begin{equation}
  \label{eq:stab-semi}
  1+2w^{(+)}_{\phi}\cos(2\beta)>0.
\end{equation}
>From Eq.~\eqref{eq:beta0-semi}
this condition requires the free twist parameter $\beta_0$
to be an increasing function of the twist parameter $\beta$.
It is derived in Appendix~\ref{sec:stab-fluct-spectr}
[see Eq.~\eqref{eq:semi-appen} with $w_{+}$ replaced by $w^{(+)}_{\phi}\cos(2\beta)$].

\subsubsection{Weak anchoring}
\label{subsubsec:weak-anch-in}

We have a somewhat different situation when the azimuthal anchoring strength
is finite at both substrates.
In this case the stability conditions can be derived using
an alternative procedure~\cite{Kis:pre:2004}.
The procedure involves two steps:
(a)~solving the linearized Euler-Lagrange equations~\eqref{eq:lin-EU};
and (b)~substituting the general solution into
the expression for the fluctuation energy~\eqref{eq:F2z}.
The last step gives the energy~\eqref{eq:F2z} expressed
in terms of the integration constants, so that the stability conditions
can be derived as conditions for this expression to be positive
definite.

Following this procedure,
we can obtain the stability conditions for the helical structures
characterized by
the parity $\mu$ and the twist parameter $\beta$
related to the free twist parameter $\beta$ through the
relation~\eqref{eq:extr2}.
The final result is
\begin{equation}
  \label{eq:stab-Hmu}
H_{\mu}=A_{\mu} + 2w^{(+)}_{\phi}w^{(-)}_{\phi}\cos v_{+}\cos v_{-}>0,
\end{equation}
\begin{equation}
  \label{eq:stab-Amu}
  A_{\mu}=\mu (w^{(+)}_{\phi}\cos v_{+}+w^{(-)}_{\phi}\cos v_{-})>0,
\end{equation}
where $v_{\pm}$ are defined in Eq.~\eqref{eq:par-vnu}.
These inequalities also
follow immediately from the stability conditions~\eqref{eq:stab-cond}
obtained in Appendix~\ref{sec:stab-fluct-spectr}
by putting $w_{\pm}=\mu w^{(\pm)}_{\phi}\cos v_{\pm}$.

Violating either of Eqs.~\eqref{eq:stab-semi}-\eqref{eq:stab-Amu}
will result in instability caused by slippage of the director
in the plane of the spiral.
Such an instability cannot occur
when the azimuthal anchoring is strong at both substrates.

\subsection{Out-of-plane fluctuations}
\label{subsec:out-of-plane}

We now study stability of the helical structures
with respect to the out-of-plane fluctuations.
To this end we replace $\lambda$ with $K_1 (2/d)^2\lambda$
and rewrite
the eigenvalue problem~\eqref{eq:gen-EV}-\eqref{eq:gen-EV-bc} for
$\theta$ in the following form:
\begin{align}
&
  \label{eq:str-EV}
  \bigl[\prt{\tau}^2-r_q/4+\lambda\bigr]\,\theta_{\lambda}(\tau) = 0,
\\
&
\label{eq:w-BC}
\Bigl[\pm\prt{\tau}
\theta_{\lambda}+w^{(\pm)}_{\theta}\theta_{\lambda}
\Bigr]_{\tau=\pm 1}=0,
\\
&
  \label{eq:rq}
  r_q=(qd)^2 K_q/K_1
=(\beta+\Delta\phi)
\bigl(
r_3 (\beta+\Delta\phi)+ 2 r_2 (\beta_0-\beta)
\bigr),
\\
&
  \label{eq:par-w-theta}
  w^{(\nu)}_{\theta}\equiv \frac{W^{(\nu)}_{\theta} d}{2K_1}
=\frac{d}{2 L^{(\nu)}_{\theta}},
\end{align}
where $\tau\equiv 2 z/d$, $r_i\equiv K_i/K_1$
and $L^{(\nu)}_{\theta}$ is the polar anchoring extrapolation length.

The stability condition $\lambda>0$ can now be readily written
as follows
\begin{equation}
  \label{eq:stab-out}
  4 \lambda_m+r_q > 0,
\end{equation}
where $\lambda_m$ is the lowest eigenvalue
of the problem~\eqref{eq:str-EV}-\eqref{eq:w-BC} computed at $r_q=0$.

When the polar anchoring is strong at both substrates, $W^{(\pm)}_{\theta}\to\infty$,
the eigenvalue $\lambda_m$ is known~[see remark at the end of Appendix~\ref{sec:stab-fluct-spectr}]:
\begin{equation}
  \label{eq:lmb-m-str}
  \lambda_m=(\kappa_m)^2=\pi^2/4.
\end{equation}
Otherwise, $\kappa_m$ is below $\pi/2$ and can be computed as the root
of  the transcendental equation deduced in
Appendix~\ref{sec:stab-fluct-spectr} [see Eq.~\eqref{eq:kappa-eq}]
\begin{equation}
  \label{eq:lambda1}
D(\kappa_m)\equiv
  (w^{(+)}_{\theta} w^{(-)}_{\theta} - \kappa_m^2)\sin 2\kappa_m+
\kappa_m (w^{(+)}_{\theta}+w^{(-)}_{\theta})\cos 2\kappa_m=0,
\end{equation}
where $0 \le\kappa_m< \pi/2$.

\subsubsection{Strong anchoring}
\label{subsubsec:strong-anchoring-out}

In the strong anchoring limit, Eq.~\eqref{eq:qd-str} implies that
the values of the twist parameter $\beta$ are quantized and do not
depend on the free twist parameter $\beta_0$.
Unstable configurations are characterized by twist wavenumbers
violating the stability condition~\eqref{eq:stab-out} with
$\lambda_m$ given in Eq.~\eqref{eq:lmb-m-str}.
These wavenumbers are described by the inequalities
\begin{align}
  \label{eq:str-unst}
&
1/\tilde{\beta}_{-}\le
  1/\tilde{\beta}\le 1/\tilde{\beta}_{+},
\quad \tilde{\beta}\equiv q d/\pi,\;
\tilde{\beta}_0\equiv q_0 d/\pi,
\notag\\
&
1/\tilde{\beta}_{\pm}=
-r_2\tilde{\beta}_0\pm
\bigl[(r_2\tilde{\beta}_0)^2+2r_2-r_3\bigr]^{1/2}.
\end{align}
These inequalities yield two different sets of unstable structures depending
on the sign of the difference $(2K_2-K_3)$.
For $q_0\ge 0$ these sets are given by
\begin{align}
  \label{eq:str-m1}
&
2 K_2 > K_3:
\notag\\
&
  q\ge (\pi/d)|\tilde{\beta}_{+}|\;\text{ or }\;
  q\le -(\pi/d)|\tilde{\beta}_{-}|,
\\
&
  \label{eq:str-m2}
K_3 > 2 K_2:
\notag\\
&
-(\pi/d)|\tilde{\beta}_{+}|\le q \le
-(\pi/d)|\tilde{\beta}_{-}|\;
\text{ at }\;
q_0 \ge \pi/(d r_2)\sqrt{r_3-2r_2}.
\end{align}

Eq.~\eqref{eq:str-m1} shows that,
when the energy cost of bend  is relatively small,
there are an infinite number of unstable configurations
and the configuration loses its stability as the distance
between its wavenumber $q$ and $q_0$ becomes sufficiently large.

Otherwise, unstable configurations may appear only if the free
wavenumber $q_0$ exceeds its critical value given in
Eq.~\eqref{eq:str-m2}. In this case the number of the unstable
configurations is finite.  From Eq.~\eqref{eq:str-m2} there is no
unstable configurations for nematic liquid crystals with $q_0=0$.
This result has been previously reported in Ref.~\cite{Gold:prl:1990}.

\subsubsection{Weak anchoring}
\label{subsubsec:weak-anchoring-out}

We now pass on to the case where
the strengths of anchoring $W^{(\pm)}_{\phi}$ are not infinitely large.
By contrast to the case of strong anchoring, the twist parameters
$\beta$ and $\beta_0$ are now not independent. Rather we have the
stationarity condition~\eqref{eq:extr2} relating
$\beta$ and $\beta_0$.
In addition, if the polar anchoring is also not infinitely strong,
the eigenvalue $\lambda_m$ can be considerably reduced.

In these circumstances,
it is reasonable to approximate the left hand side of the stability
condition~\eqref{eq:stab-out} by its lower bound derived in the limit
of weak polar anchoring, $W^{(\nu)}_{\theta}\to 0$, where
$\lambda_m$ vanish. Technically, the resulting condition
\begin{equation}
  \label{eq:th-cnd}
  r_q(\beta,\mu)=
(\beta+\Delta\phi)
\bigl(\,
2\, r_2\, \beta_0(\beta,\mu)+
(r_3 -2 r_2)\beta+r_3 \Delta\phi\,
\bigr) > 0
\end{equation}
is sufficient but not necessary for stability.
Thus, when the inequality~\eqref{eq:th-cnd} is satisfied,
the structure will certainly be locally stable with respect to out-of-plane fluctuations
whatever the polar anchoring is.

Eq.~\eqref{eq:th-cnd} can now be used to study
out-of-plane fluctuation induced instability of
the helical structures which are otherwise stable
with respect to in-plane fluctuations and thus meet the stability
conditions~\eqref{eq:stab-Hmu}-\eqref{eq:stab-Amu}.
For $\Delta\phi=0$ and positive twist wavenumbers, such instabilty
may occur only if the doubled twist elastic constant exceeds the bend
elastic constant, $2K_2>K_3$, and the azimuthal anchoring energy is
sufficiently large. In this case, however, $r_q$ can be made
non-negative by  increasing the value of the free twist parameter
$\beta_0$.  In other words, if the ratio of the cell thickness and
the equilibrium CLC pitch is large enough to meet the
condition~\eqref{eq:th-cnd} we can neglect out-of-plane deviations of the director and use
the ``smectic-like'' free energy~\eqref{eq:en-n0}.

\subsection{Correlation functions}
\label{subsec:corr-funct}

Our calculations of
the director fluctuation static correlation function $\langle\bs{\psi}(z)\bs{\psi}(z')\rangle$
use the relation~\cite{Kis:pre:2004}
\begin{equation}
  \label{eq:corr-fun}
  \langle\bs{\psi}(z)\bs{\psi}(z')\rangle=
\frac{k_B T}{S}\,\vc{G}(z,z'),
\end{equation}
where $k_B$ is the Boltzmann constant and
$T$ is the temperature.
The Green function  $\vc{G}(z,z')$ can be computed as the inverse
of the operator $\hat{K}$ defined in Eq.~\eqref{eq:op-K}  by solving the boundary-value problem
\begin{align}
  \label{eq:gen-corr}
  \hat{K}\vc{G}(z,z')=\delta(z-z')\hat{I},
\\
\label{eq:gen-corr-bc}
\hat{Q}^{(\nu)}
\vc{G}(z,z') \Bigr\vert_{z=\nu d/2}=0,
\end{align}
where $\hat{I}$ is the identity matrix and the operator
$\hat{Q}^{(\nu)}$ is given in Eq.~\eqref{eq:op-Q}.
Since the matrix operators $\hat{K}$ and $\hat{Q}^{(\nu)}$ are both diagonal,
the correlation function~\eqref{eq:corr-fun} is also diagonal:
\begin{equation}
  \label{eq:corr-fun-diag}
  \langle\bs{\psi}(z)\bs{\psi}(z')\rangle=
\begin{pmatrix}
\langle\phi(z)\phi(z')\rangle & 0\\
0 & \langle\theta(z)\theta(z')\rangle
\end{pmatrix}.
\end{equation}

Solving the boundary-value
problem~\eqref{eq:gen-corr}-\eqref{eq:gen-corr-bc}
yields the following expressions for
the in-plane  and the out-of-plane components of
the correlation function~\eqref{eq:corr-fun-diag}
\begin{align}
&
  \label{eq:th-corr}
  \langle\theta(z)\theta(z')\rangle=
\frac{k_B T d}{ 2S K_1\kappa D(\kappa)}\,
\Bigl[
w^{(-)}_{\theta}\sin\kappa(1+2 z_{<}/d)
\notag\\
&
+
\kappa\cos\kappa(1+2 z_{<}/d)
\Bigr]
\Bigl[
w^{(+)}_{\theta}\sin\kappa(1-2 z_{>}/d)
+
\kappa \cos\kappa(1-2 z_{>}/d)
\Bigr],
\end{align}
\begin{align}
&
 \label{eq:phi-corr}
  \langle\phi(z)\phi(z')\rangle =
\frac{k_B T d}{ 2 S K_2 H_{\mu}}\,
 \Bigl[ 1
 + \mu (1+2 z_{<}/d) w^{(-)}_{\phi}\cos v_{-} \Bigr]
\notag
\\
&
\times
 \Bigl[ 1
 + \mu (1-2 z_{>}/d) w^{(+)}_{\phi}\cos v_{+} \Bigr],
\end{align}
where $\kappa^2=-r_q/4$, $z_{<}\equiv\min\{z,z'\}$,
$z_{>}\equiv\max\{z,z'\}$,
$D(\kappa)$ and $H_{\mu}$ are defined in Eq.~\eqref{eq:lambda1}
and Eq.~\eqref{eq:stab-Hmu}, respectively.
The correlation functions diverge on approaching the boundary
of  the stability region. For out-of-plane fluctuations, the
denominator of the expression for $\langle\theta(z)\theta(z')\rangle$
vanish in the limit of marginal stability
where $-r_q/4\to\lambda_m=\kappa_m$ and $D(\kappa)\to D(\kappa_m)=0$. Similarly,
Eq.~\eqref{eq:stab-Hmu} shows that, in the marginal stability
limit for in-plane fluctuations,  $H_{\mu}$ goes to zero  thus rendering
the correlation function $\langle\phi(z)\phi(z')\rangle$ divergent.

\begin{figure*}[!tbh]
\centering
\resizebox{150mm}{!}{\includegraphics*{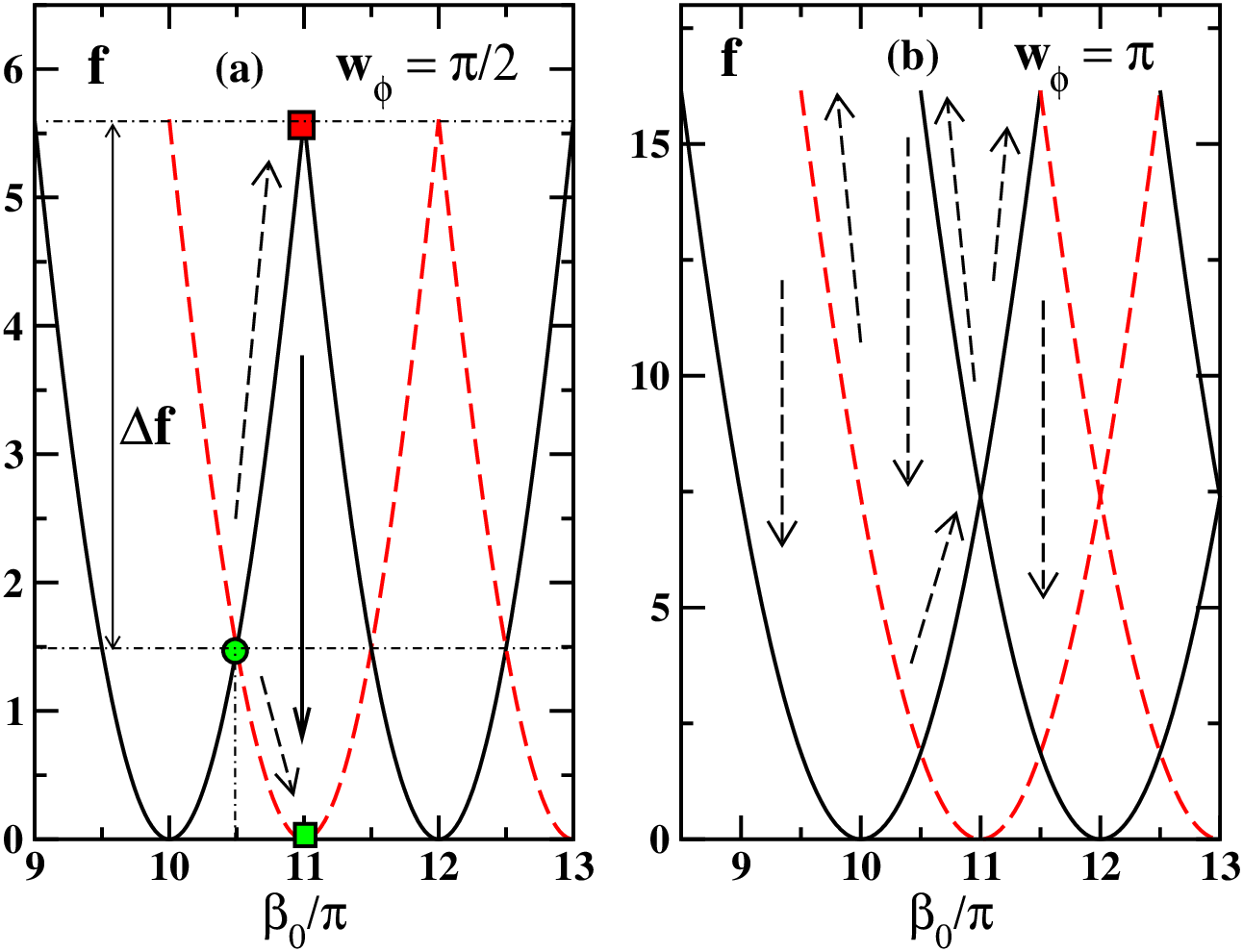}}
\caption{%
The free energy $f_{+}$ (solid line) and $f_{-}$ (dashed line)
of stable configurations with the half-turn number, $k$,
between $10$ and $14$
as a function of $\beta_0$ computed from Eq.~\eqref{eq:fmu}
by using Eq.~\eqref{eq:extr2}
for two values of the anchoring energy parameters:
(a)~$w_{\phi}=\pi/2$; and (b)~$w_{\phi}=\pi$.
The intersection point of the branches with
$k=10$ and $k=11$ at $\beta_0=10.5\pi$ is indicated by circle.
Squares mark
energy of the structures at the transition point with $\beta_0=11\pi$
where the configuration with $k=10$ loses its stability.
It is shown that the energy of the helical
structure with $k=10$ ($k=11$) increases (decreases) as $\beta_0$ varies
from $10.5\pi$ to $11\pi$.
}
\label{fig:en-bet0}
\end{figure*}

\begin{figure*}[!tbh]
\centering
\resizebox{150mm}{!}{\includegraphics*{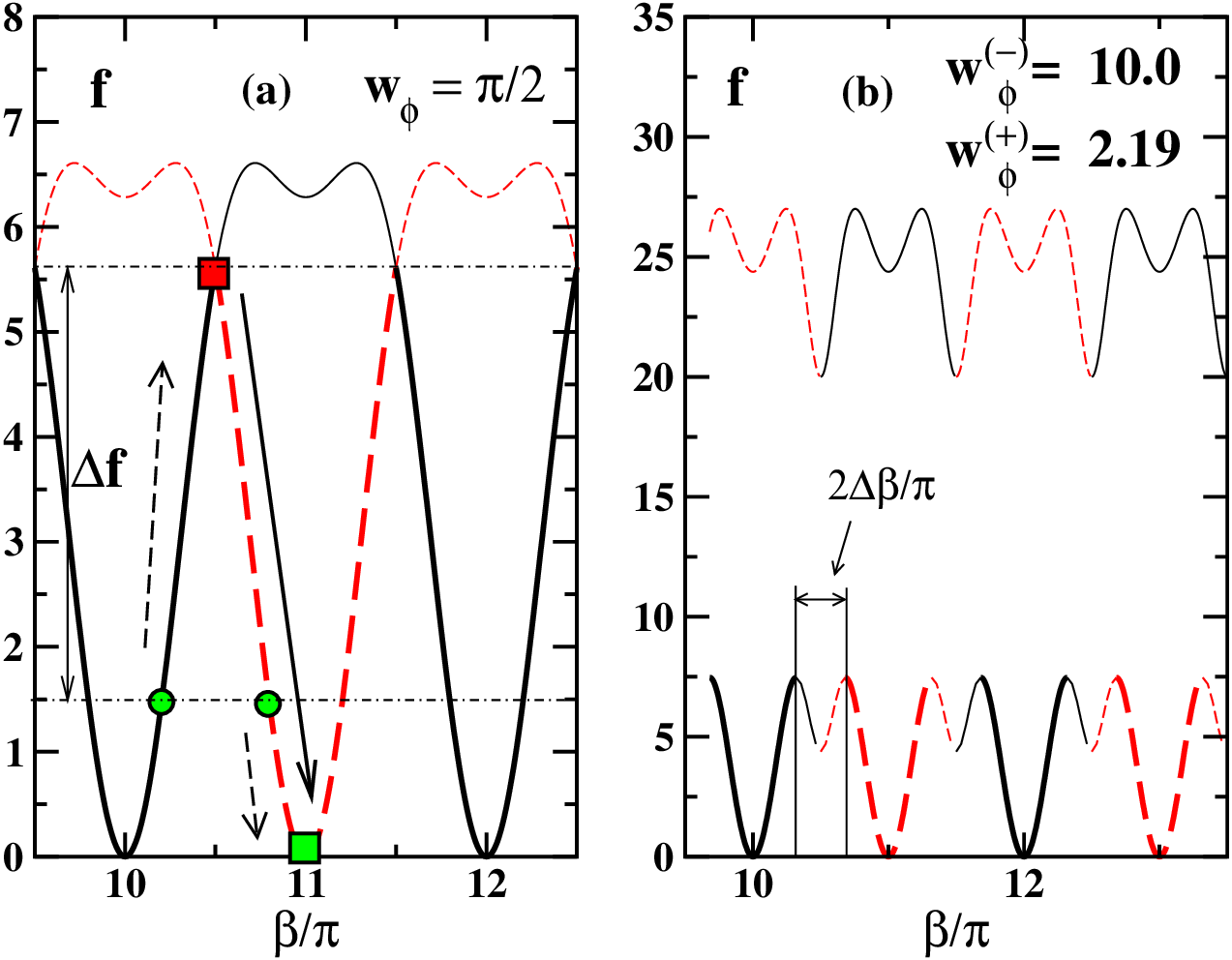}}
\caption{%
The free energy $f_{+}$ (solid line) and $f_{-}$ (dashed line)
of the configurations with the half-turn number, $k$,
between $10$ and $14$
as a function of $\beta$ computed from Eq.~\eqref{eq:fmu}.
Thin lines represent the energy of unstable configurations.
Two cases are shown:
(a) $w^{(-)}_{\phi}=w^{(+)}_{\phi}=w_{\phi}=\pi/2$; and
(b) $w^{(-)}_{\phi}=10.0$, $w^{(+)}_{\phi}=2.19$
 ($\sigma=5.6$ and $w_{\phi}=\pi/2$).
Circles and squares label energy of the structures at $\beta_0=10.5\pi$
and $\beta_0=11\pi$ shown in Fig.~\ref{fig:en-bet0}(a).
At $\beta_0=10.5\pi$, the structures with $k=10$ and $k=11$
are shown to be degenerate in energy and separated by the energy
barrier $\Delta f$.
}
\label{fig:en-bet}
\end{figure*}

\begin{figure*}[!tbh]
\centering
\resizebox{150mm}{!}{\includegraphics*{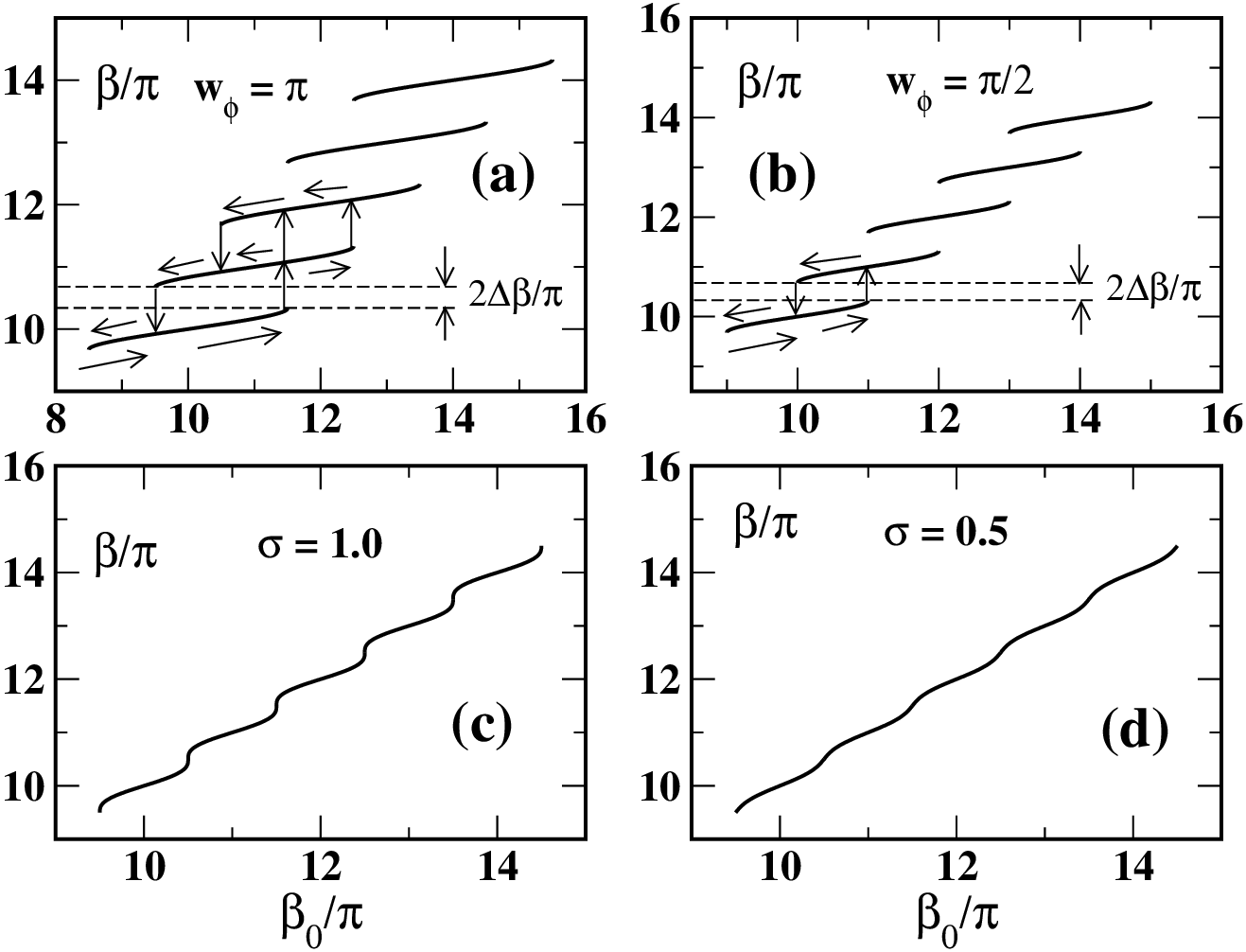}}
\caption{%
Dependence of $\beta$  on $\beta_0$
calculated at $w_{-}=10.0$
for various values of the parameters
$\sigma$ ($=2w^{(-)}_{\phi}w^{(+)}_{\phi}/(w^{(-)}_{\phi}-w^{(+)}_{\phi})$) and $w_{\phi}$ (see Eq.~\eqref{eq:ww}):
(a) $w_{\phi}=\pi$ ($\sigma= 11.8$ and $w^{(+)}_{\phi}= 3.705$);
(b) $w_{\phi}=\pi/2$ ($\sigma= 5.6$ and $w^{(+)}_{\phi}= 2.19$);
(c) $\sigma=1.0$ ($w^{(+)}_{\phi}=w_c= 0.476$);
(d) $\sigma=0.5$ ($w^{(+)}_{\phi}= 0.25$).
}
\label{fig:asym}
\end{figure*}

\begin{figure*}[!tbh]
\centering
\resizebox{150mm}{!}{\includegraphics*{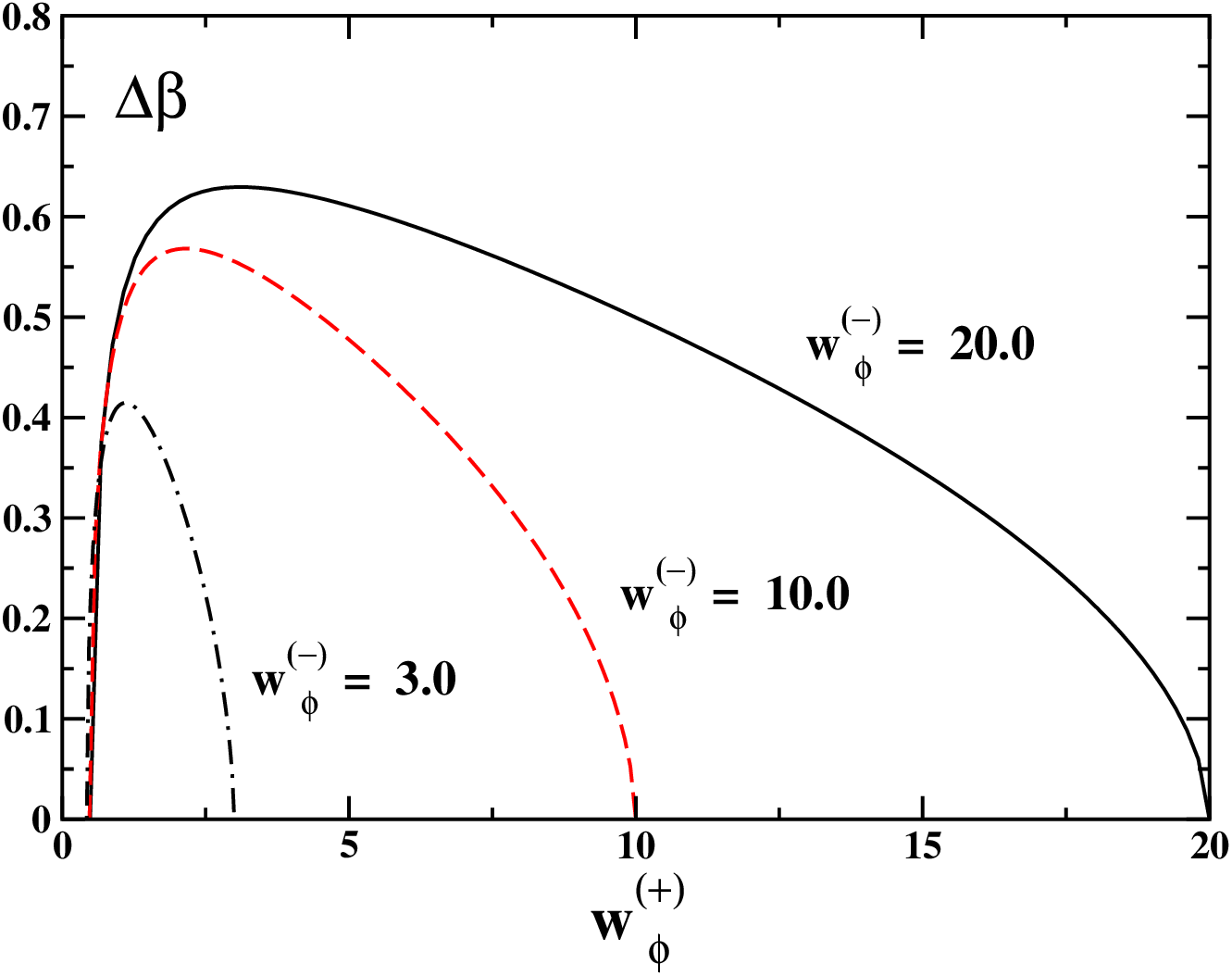}}
\caption{%
Dependence of $\Delta\beta$  on $w^{(+)}_{\phi}$
for various values of the anchoring energy parameter
$w^{(-)}_{\phi}$.
}
\label{fig:delt-w}
\end{figure*}

\begin{figure*}[!tbh]
\centering
\resizebox{150mm}{!}{\includegraphics*{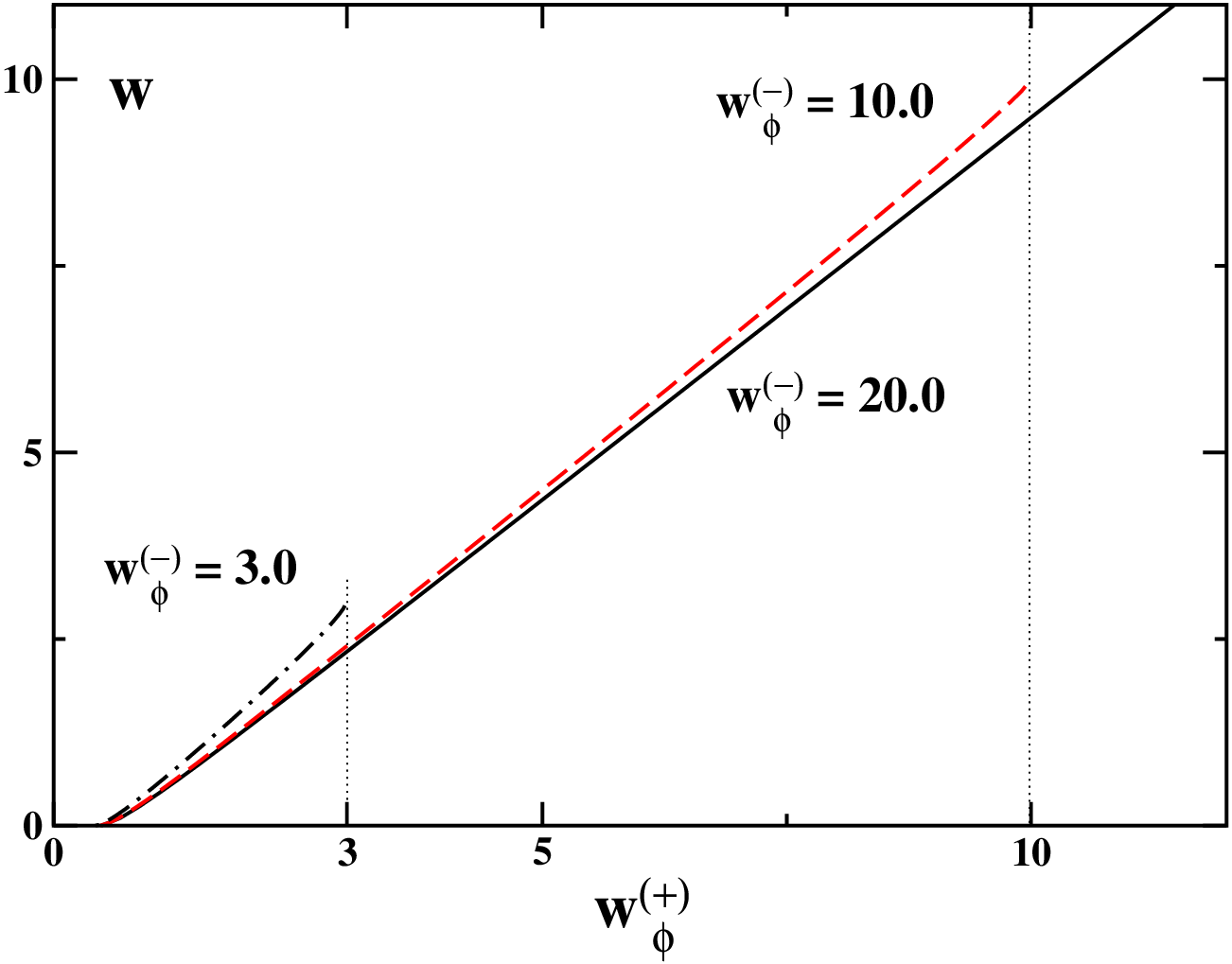}}
\caption{%
Dependence of $w_{\phi}$  on $w^{(+)}_{\phi}$
for various values of the anchoring energy parameter
$w^{(-)}_{\phi}$. It is shown that $w_{\phi}=w^{(-)}_{\phi}$ at $w^{(+)}_{\phi}=w^{(-)}_{\phi}$
and $w_{\phi}=0$ at $w^{(+)}_{\phi}=w_c$ (see Eq.~\eqref{eq:w_crit}).
}
\label{fig:ww-w}
\end{figure*}

\begin{figure*}[!tbh]
\centering
\resizebox{150mm}{!}{\includegraphics*{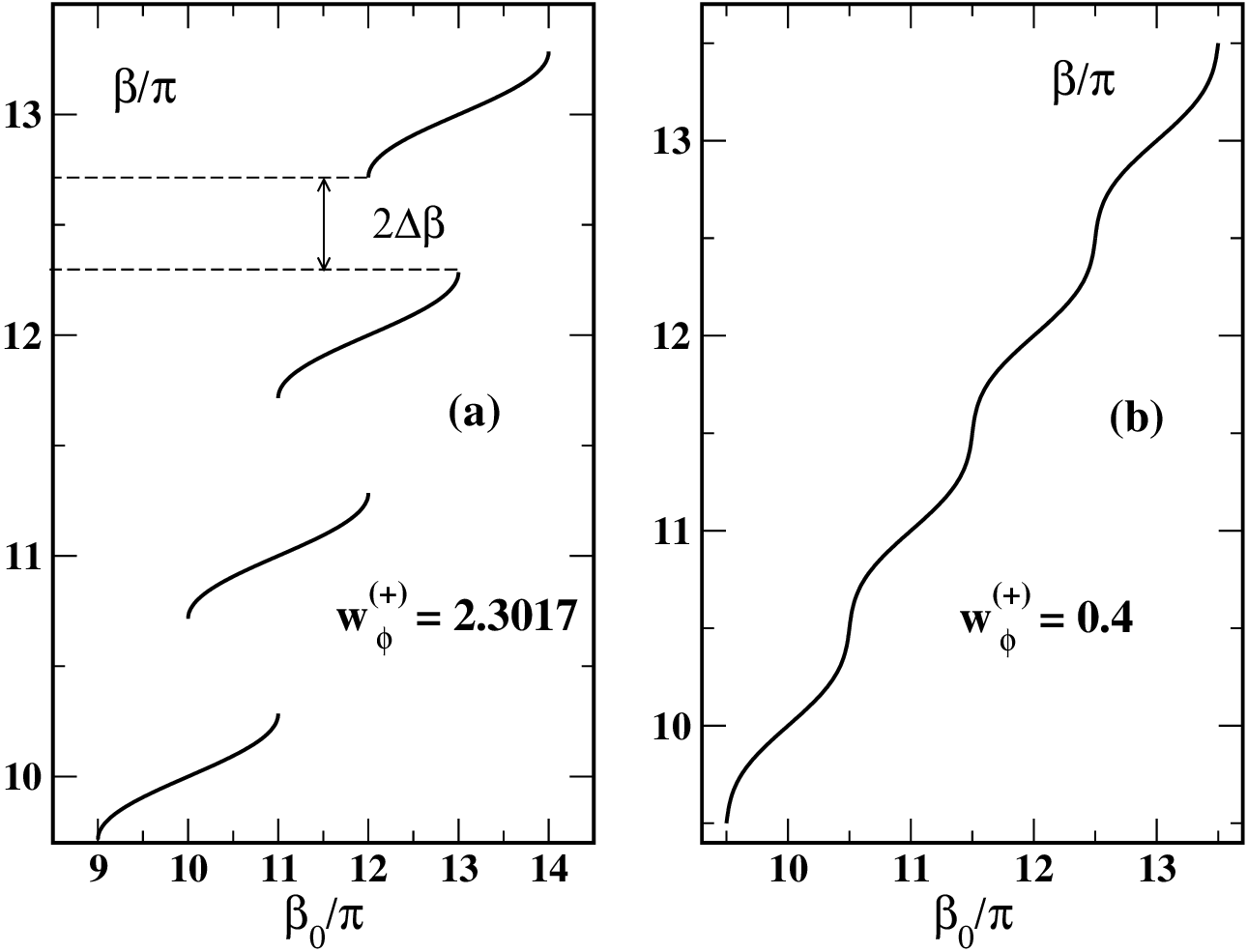}}
\caption{%
Twist parameter $\beta$  as a function of $\beta_0$
in the limit of semi-strong anchoring, $w^{(-)}_{\phi}\to\infty$.
Two cases are shown:
(a)~$w^{(+)}_{\phi}= 2.3017>0.5$ with $w_{\phi}=\pi/2$
(see Eq.~\eqref{eq:ww-semi});
and
(b)~$w^{(+)}_{\phi}= 0.4<0.5$.
}
\label{fig:semi}
\end{figure*}

\section{Transitions induced by free wavenumber variations}
\label{sec:pitch-vs-twist}

In the previous section we have studied
the stability of the CLC helical structures~\eqref{eq:n0}
with respect to both  in-plane and out-of-plane fluctuations.
We have found that the anchoring conditions play a crucial role in the
calculations. In particular, cells with strong anchoring and those
with what we have called semi-strong anchoring exhibit significantly
different properties.

In this section we concentrate on the weak anchoring cases.
We have shown that in this case helical structures are characterized by
the twist parameter, $\beta$, and the half-turn parity, $\mu$.
These quantities are related to the free twist
parameter, $\beta_0$, through the stationary point equation~\eqref{eq:extr2}.
The structure responds to variations of the free wavenumber
(and thus the free twist parameter) by changing its twist parameter.

This change may render the initially equilibrium structure either
metastable or unstable.
When the anchoring is not infinitely strong and the free twist
parameter is large enough to meet the stability
condition~\eqref{eq:th-cnd},
this instability is solely governed by in-plane director fluctuations
and defines the mechanism dominating
transformations of the director field.
This mechanism is suppressed in the strong anchoring regime,
where the structural transitions involve tilted
configurations~\cite{Gold:prl:1990},
and  can be described as
director slippage through the energy barriers formed by the surface
potentials.

In this section our task is to study helical structure transformations
as a function of
the free twist wavenumber $q_0$ for different anchoring conditions.
Equivalently, we focus our attention on the dependence of $\beta$
on $\beta_0$; this can be thought of as a sort
of dispersion relation. To this end we examine in more detail
the consequences of the analytical
results obtained in the previous sections, Sec.~\ref{sec:general}
and Sec.~\ref{sec:stability}.

\subsection{Symmetric cells}
\label{subsec:equal-anch}

When the anchoring strengths at both  substrates are equal,
$W^{(-)}_{\phi}=W^{(+)}_{\phi}\equiv W_{\phi}$, the right hand side of Eq.~\eqref{eq:extr2}
is  $\beta\pm w_{\phi}\sin\beta$ ($w_{\phi}=W_{\phi}d/2K_2$)
and $v_{\pm}=\beta$.
In this case stability of the configurations is governed
by Eq.~\eqref{eq:stab-Amu}
which reduces to the simple inequality $\mu\cos\beta >0$.

It immediately follows that
the values of  $\beta$
representing the locally stable structures of the parity $\mu$
ranged  between
$(k-1/2)\pi$ and $(k+1/2)\pi$, where $k$ is the even (odd) integer
at $\mu=+1$ ($\mu=-1$). The integer $k$ will be referred to as
the half-turn number. The parity $\mu$
introduced in Sec.~\ref{sec:general} is now shown
to be the parity of the half-turn number: $\mu=(-1)^k$.

The intervals of $\beta$ for the stable configurations
$[(k-1/2)\pi,(k+1/2)\pi]$ are
now labelled by the half-turn number $k$.
Since the function $\gamma_{\mu}$ ($\mu=(-1)^k$) monotonically increases
on the interval characterized by the half-turn number $k$,
the value of $\beta_0$ runs from
$(k-1/2)\pi-w_{\phi}$ to $(k+1/2)\pi+w_{\phi}$ on this interval.
As a result, for each half-turn number $k$,
there is the monotonically increasing
branch of the $\beta(\beta_0)$ curve.

The branches with $k$ ranged from $10$ to $13$ for
different values of the dimensionless anchoring energy parameter
$w_{\phi}$ are depicted in Figs.~\ref{fig:sym}(b)-(d).
We see that the $\beta_0$-dependence of $\beta$ will
always be discontinuous provided the anchoring energy is not equal to zero.
Fig.~\ref{fig:sym}(d) shows that the jumps tend to disappear
in the limit of weak anchoring, where the azimuthal anchoring energy
approaches zero, $w_{\phi}\to 0$.

As we pointed out in Sec.~\ref{subsubsec:strong-anch-limit}
for the case of strong anchoring,
there are two equilibrium structures of the same energy
at $\beta_0=(1/2+m)\pi$. In Fig.~\ref{fig:sym}(a)
the arrows indicate that the half-turn number of the equilibrium
structure changes at these points.

Similarly,
when the anchoring is weak, $w_{\phi}<\pi$, and $\beta_0=(1/2+m)\pi$,
Eq.~\eqref{eq:extr2} possess two different roots
with $k=m$ and $k=m+1$
which are equally distant from $\beta_0$ and are of equal energy.
In Fig.~\ref{fig:en-bet0}(a), the free
energy~\eqref{eq:fmu} is shown as a function of $\beta_0$.
It can be seen that the intersection points of
the curves for different parities,
(solid and dashed lines in Fig.~\ref{fig:en-bet0}) are indeed at
$\beta_0=(1/2+m)\pi$.
The parity of the equilibrium configuration reverses
as $\beta_0$ goes through the values $(1/2+m)\pi$.
Fig.~\ref{fig:en-bet0}(b) illustrates that this is also the case even
if $w_{\phi}\ge\pi$.

For $m=10$, Fig.~\ref{fig:en-bet0}(a) and Fig.~\ref{fig:en-bet}(a)
show the initially equilibrium structure with the half-turn number
$k=10$ (solid line) becomes metastable
as $\beta_0$ passes through the critical point $\beta_0=(1/2+m)\pi$
at which the structures with $k=10$ and $k=11$ (dashed line) are
degenerate in energy.
In Fig.~\ref{fig:en-bet0}(a) and Fig.~\ref{fig:en-bet}(a)
the structures at this point are indicated by circles.

As is seen from the figures,
relaxation to the new equilibrium state will require
the jump-like change of the twist parameter $\beta$.
In addition, the transition between the metastable and the equilibrium
configurations involves penetrating the energy barrier $\Delta f$
that separates the states with different half-turn numbers.
This barrier can be seen from Figs.~\ref{fig:en-bet0}(a)
and~\ref{fig:en-bet}(a) where  the free energy~\eqref{eq:fmu}
is plotted as a function of $\beta_0$ and $\beta$, respectively.

Previous authors~\cite{Zink:1999,Bel:eng:2000,Palto:eng:2002}
have supposed that the transitions may occur only if there is no energy barrier.
Clearly, this assumption implies that the jumps take place at
the end points of the stability intervals:
$\beta_0=\beta_{\pm}^{(k)}=(1/2+k)\pi \pm w_{\phi}$,
where the configuration
with the half-turn number $k$ becomes marginally stable
($ A_{\mu}=H_{\mu}=0$) and loses its stability.

These transitions are indicated by arrows in
Figs.~\ref{fig:sym},~\ref{fig:en-bet0} and~\ref{fig:asym}.
As is seen from Fig.~\ref{fig:sym}(c),
in this case the upward and backward
transitions: $k\to k+1$ and $k+1\to k$ occur
at different values of $\beta_0$:
$\beta_{+}^{(k)}$ and $\beta_{-}^{(k+1)}=\beta_{+}^{(k)} - 2w_{\phi}$,
respectively. So, there are hysteresis loops in
the response of CLC cell to the change in the free twisting number.

We can now describe how the increase in the anchoring energy
will affect the scheme of the transitions.
To be specific, we consider the critical end point
$\beta_0=\beta_{+}^{(k)}$, so that
for small anchoring energies with $w_{\phi} < \pi/2$
there are only two configurations:
the marginally stable initial configuration
with $\beta=\beta_k$ and the equilibrium structure with
$\beta=\beta_{k+1}$. In this case
Eq.~\eqref{eq:extr2} has at most two roots and
the jumps will occur as transitions between the states
which  half-turn numbers differ by the unity, $|\Delta k|=1$.

At $w_{\phi}=\pi/2$, as shown in
Fig.~\ref{fig:en-bet0}(a),
we have two marginally stable structures of equal energy:
$\beta_{k}$ and $\beta_{k+2}$.
The newly formed structure $\beta_{k+2}$
being metastable at $\pi/2< w_{\phi}<\pi$  will have the free energy
equal to the energy of the equilibrium configuration $\beta_{k+1}$
at $w_{\phi}=\pi$.
So, as illustrated in Figs.~\ref{fig:sym}(b) and~\ref{fig:en-bet0}(b), both
transitions $k\to k+1$ and $k\to k+2$ are equiprobable
and we have the bistability  effect at the critical point
under $w_{\phi}=\pi$.

For $\pi< w_{\phi} <3\pi/2$ there are three configurations:
the initial configuration $\beta_k$,
the metastable configuration $\beta_{k+1}$
and the equilibrium structure $\beta_{k+2}$.
The configuration $\beta_{k+3}$ being formed at $w_{\phi}=3\pi/2$
will define the equilibrium structure at $2\pi \le w_{\phi}\le 3\pi$
and so on.

The general result for the critical point $\beta_0=\beta_{+}^{(k)}$
can be summarized as follows.
When $(l+1/2)\pi < w_{\phi} < (l+3/2)\pi$, in addition to the marginally
stable configuration $\beta_k$,
there are $l+2$ stable configurations with the half-turn numbers
ranged from $k+1$ to $k+l+2$. The half-turn number of the equilibrium
structure equals $k+l+1$ under $l\pi < w_{\phi} < (l+1)\pi$.

It immediately follows that the restriction $w_{\phi}< \pi$
imposed by Belyakov and Kats~\cite{Bel:eng:2000} on the anchoring strength
requires the relaxation transitions to involve only two structures
with $|\Delta k|=1$. Our result shows that, when the anchoring
parameter $w_{\phi}$ falls between $l\pi$ and $(l+1)\pi$,
the half-turn number change is $|\Delta k|=l+1$ for
the transitions between marginally stable and the equilibrium states.
Clearly, we can have the transitions with even $\Delta k$ that
involve topologically equivalent configurations with common parity~\cite{Klem:rpp:1989,Oswald:prep:2000}.
Such transitions may also be induced by the thermal director fluctuations
without formation of defects even if the anchoring is
infinitely strong~\cite{Gold:prl:1990}.
Though the mechanism under consideration is rather different,
neglecting the director fluctuations
can only be regarded as a zero-order approximation.

Indeed, according to our remark at the end of Sec.~\ref{subsec:out-of-plane}, the expression for
the fluctuation  correlation function~\eqref{eq:phi-corr}
implies its divergence upon reaching a marginally stable
state where $H_{\mu}=0$. It means that
taking the fluctuations into account will give
the transition points located within the stability interval.
This fluctuation induced shift may also suppress the hysteresis
provided the mean square angle deviation
$\sqrt{\langle\phi^2\rangle}$, computed from Eq.~\eqref{eq:phi-corr}
at $\beta_0=(1/2+m)\pi$ and $z=z'=d/2$, and the anchoring
energy parameter $w_{\phi}$ are of the same order.

\subsection{Asymmetric cells}
\label{subsec:diff-anch}

When the anchoring energies at the surfaces are different,
$W^{(-)}_{\phi}\ne W^{(+)}_{\phi}$ and $\epsilon\ne 0$,
$\sin v_{+}$ on the right hand side of Eq.~\eqref{eq:extr2}
equals zero at $\beta=\pi/2+\pi k$
and, as demonstrated in Fig.~\ref{fig:inter}(b), we have additional
intersection points of the curves $\gamma_{+}(\beta)$
and $\gamma_{-}(\beta)$. It can be shown that
the stability conditions are now defined by Eq.~\eqref{eq:stab-Hmu}
and the twist parameters  $\beta$ of
the marginally stable configurations, where
$H_{\mu}=0$,
can be computed as the stationary points of $\gamma_{\mu}$.

These points represent
the local maxima and minima of $\gamma_{\mu}$ and are located at
$\beta=(1/2+k)\pi\pm \Delta\beta$.
The equation for $\Delta\beta$ is
\begin{equation}
  \label{eq:delt-bet}
  w^{(+)}_{\phi}(1+\epsilon) \sin\bigl(\Delta\beta
-\epsilon
\arctan[
\epsilon \cot\Delta\beta]
\bigr)=
-\frac{1+(\epsilon^2-1)\cos^2\Delta\beta}{1+(\epsilon-1)\cos^2\Delta\beta}\,,
\end{equation}
where $\Delta\beta\in [0,\pi/2]$.

>From the stability condition $H_{\mu}>0$
the values of
$\beta$ for stable configurations fall  between
the stationary points
$(k-1/2)\pi+\Delta\beta$ and $(k+1/2)\pi-\Delta\beta$,
where the half-turn number $k$ is the even (odd) integer
depending on the parity.
The function $\gamma_{\mu}$ monotonically increases
and $\beta_0$ varies from
$(k-1/2)\pi-w_{\phi}$ to $(k+1/2)\pi+w_{\phi}$ on the stability interval
with the half-turn number $k$.
The effective dimensionless anchoring parameter $w_{\phi}$,
as opposed to the case of equal anchoring energies with $w^{(\pm)}_{\phi}=w_{\phi}$,
is now given by
\begin{equation}
  \label{eq:ww}
  w_{\phi}=w^{(+)}_{\phi}\cos\bigl(\Delta\beta
-\epsilon
\arctan[
\epsilon \cot\Delta\beta]
\bigr)-\Delta\beta\,.
\end{equation}

Clearly, we can now  follow the line of reasoning
presented in Sec.~\ref{subsec:equal-anch} to find out
the results concerning hysteresis loops and bistability
effects that are quite similar to the case of equal anchoring
strengths (see Figs.~\ref{fig:asym}(a)-(b)).
There are, however, two important differences related
to Eqs.~\eqref{eq:delt-bet} and~\eqref{eq:ww}.

If $\Delta\beta\ne 0$,
the intervals of $\beta$ representing stable director configurations
are separated by the gap of the length $2\Delta\beta$.
The presence of this gap is illustrated in Figs.~\ref{fig:asym}(a)
and~\ref{fig:asym}(b). Fig.~\ref{fig:en-bet}(b) shows the gap
between stable branches of the dependence of the free energy
on $\beta$.
The values of $\beta$  within the gap represent
unstable configurations and form the zone of
``forbidden'' states in the CLC cell.

The graph of the $\Delta\beta$ vs $w^{(+)}_{\phi}$
dependence is presented in Fig.~\ref{fig:delt-w}.
As expected, the gap is shown to
disappear in the limit of equal energies, $w^{(+)}_{\phi}=w^{(-)}_{\phi}$.
Another and somewhat more interesting effect is
that there is a small critical
value of $w^{(+)}_{\phi}$ below which $\Delta\beta$ also vanishes.

In order to interpret this effect, we note
that Eq.~\eqref{eq:extr2} with $\beta_0=(1/2+k)\pi$
has the only solution, $\beta=\beta_0$,
provided the azimuthal anchoring energy parameters meet the condition:
\begin{equation}
\sigma\equiv\frac{2w^{(-)}_{\phi}w^{(+)}_{\phi}}{|w^{(-)}_{\phi}-w^{(+)}_{\phi}|}\le 1\,.
  \label{eq:crit}
\end{equation}
Another form of this condition
\begin{equation}
  \label{eq:crit2}
  |L^{(+)}_{\phi} - L^{(-)}_{\phi}|\ge d
\end{equation}
implies that the difference between the azimuthal anchoring
extrapolation lengths is larger than the cell thickness.
For hybrid cells, similar inequality is known as the stability
condition of homogeneous structures~\cite{Babri:jpf:1983,Spar:pre:1994,Zih:pre:2000}.

In this case the gap disappears and the dependence of
$\beta$ on $\beta_0$ becomes continuous in the manner
indicated in Figs.~\ref{fig:asym}(c) and~\ref{fig:asym}(d).
Given the value of $w^{(-)}_{\phi}$ the relation~\eqref{eq:crit}
yields the threshold value for
the anchoring strength at the upper substrate:
\begin{equation}
w_c=\frac{w^{(-)}_{\phi}}{2 w^{(-)}_{\phi}+1}\,.
  \label{eq:w_crit}
\end{equation}

So, the jumps and the gap will vanish at $w^{(+)}_{\phi}=w_c$.
Analogously, as illustrated in Fig.~\ref{fig:ww-w},
$w_{\phi}$ goes to zero at the critical point $w^{(+)}_{\phi}=w_c$,
while for large values of $w^{(-)}_{\phi}$
the dependence of $w_{\phi}$ on $w^{(+)}_{\phi}$ is approximately linear.

In closing this section we discuss the limiting case of semi-strong
anchoring where  $w^{(-)}_{\phi}\to\infty$.
For this purpose we can combine the stabillity
condition~\eqref{eq:stab-semi}
with Eq.~\eqref{eq:beta0-semi} linking
the free twisting parameter $\beta_0$ and the twist parameter $\beta$
of the helical structures chararcterized by the
energy~\eqref{eq:fmu-semi}.

>From the stabillity condition~\eqref{eq:stab-semi} the gap separating the
stability intervals ranged between $(k-1/2)\pi+\Delta\beta$
and $(k+1/2)\pi-\Delta\beta$ is given by
\begin{equation}
  \label{eq:delt-bet-semi}
  2\Delta\beta=
\begin{cases}
\arccos \bigr(2 w^{(+)}_{\phi}\bigl)^{-1}, & 2 w^{(+)}_{\phi}>1,\\
0, & 2 w^{(+)}_{\phi}\le 1.
\end{cases}
\end{equation}
This result also follows from Eq.~\eqref{eq:delt-bet} in
the semi-strong limit $\epsilon\to 1$.
Similarly, the expression for $w_{\phi}$~\eqref{eq:ww} simplifies to
the following form
\begin{equation}
  \label{eq:ww-semi}
  w_{\phi}=w^{(+)}_{\phi}\sin\bigl(2\Delta\beta\bigr)-\Delta\beta\,.
\end{equation}

Eqs.~\eqref{eq:delt-bet-semi} and~\eqref{eq:ww-semi} explicitly show
that the gap and the hysteresis loops both disappear when the
anchoring strength is sufficiently small and
$2 w^{(+)}_{\phi}\le 1$.
In other words, when the inequality
\begin{equation}
  \label{eq:crit-semi}
  L^{(+)}_{\phi} \ge d
\end{equation}
is satisfied, the discontinuities turn out to be suppressed.

Thus, in asymmetric CLC cells, the left hand side of
Eqs.~\eqref{eq:crit2} and~\eqref{eq:crit-semi}
define the critical cell thickness below which the $\beta$ vs
$\beta_0$ curve becomes continuous.
For semi-strong anchoring, this effect can be seen from the curves
shown in Fig.~\ref{fig:semi}.

\section{Discussion and conclusions}
\label{sec:concl}

In this paper we have studied how the pitch wavenumber $q$
of the helical director configuration
in the CLC cell depends on the free twisting number $q_0$
at different anchoring conditions imposed by
the cell substrates. It is found that
this dependence is generally discontinuous
and is characterized by the presence of hysteresis and bistability.

We have shown that asymmetry in
the strengths of the director anchoring with the substrates
introduces the following new effects:
\renewcommand{\theenumi}{\alph{enumi}}
\renewcommand{\labelenumi}{(\theenumi)}
\begin{enumerate}
\item
the jump-like behaviour of the twist wavenumber is suppressed
when the difference between the azimuthal anchoring extrapolation
lengths is larger than the cell thickness;

\item
the twist wavenumber intervals of locally stable configurations
with adjacent numbers of helix half-turns are separated by
gap in which the structures are unstable.
\end{enumerate}

Using stability analysis we have emphasized the idea that
the instability mechanism
behind the transitions between the helical structures is
dominated by the in-plane director fluctuations.
These fluctuations may render the
structures unstable only if the anchoring energy is finite.

In this case the height of the energy barriers separating the CLC
states in the space of in-plane variables is determined by the surface
potentials and is also finite. The mechanism can, therefore, be described as
slippage of the director through the anchoring energy barrier.
Interestingly, a similar mechanism can be expected be important
in trying to extend the theory of Ref.~\cite{Sluc:pre:2000},
where shear-induced melting of smectic-$A$ liquid crystals
has been studied in the strong anchoring limit,
to the case of weak anchoring.

The part of our analysis presented in Sec.~\ref{subsec:equal-anch}
relies on the assumption that
the transition between configurations with different
half-turn numbers occurs when the initial structure loses its
stability, so that pitch wavenumber is no longer
a local minimum of the free energy surface.
The result is that
the stronger the anchoring, the larger the change of the half-turn
number (and of the twist wavenumber) needed to reach the equilibrium
state. So, whichever mechanism of relaxation is assumed,
the metastable states certainly play an important part in the problem
when the anchoring is not too weak.

It was recently shown by Bisi \textit{et. al}~\cite{Bisi:pre:2003} that
the instability mechanism in twisted nematics may involve
the so-called eigenvalue exchange
configurations~\cite{TJS:prl:1987,Palf:lc:1994}.
These configurations and the tilted structures are, however, of minor importance
for the director slippage induced instability. They may be important
outside the parameter regime considered here, and
 we will discuss alternative mechanisms in more detail elsewhere.

The dynamics of the transitions is well beyond the scope of this paper.
Despite some very recent progress~\cite{Bel:eng:2004},
it still remains a challenge to develop a tractable theory
that properly account for director fluctuations, hydrodynamic modes
and defect formation.
Simultaneously we have seen at the end of Sec.~\ref{subsec:equal-anch} that
fluctuation effects can be estimated by using
the expression for the correlation functions given in
Sec.~\ref{subsec:corr-funct}.
But in order to take the fluctuations into consideration
a systematic treatment is required.

 \begin{acknowledgments}
This work was partially carried out in the framework of a UK-Ukraine
joint project funded by the Royal Society. T.J.S.  is grateful to
V.A. Belyakov and E.I. Kats for useful conversations, correspondence
and for sending copies of preprints of relevant papers.
A.D.K. thanks the School of Mathematics in the University of Southampton
for hospitality during his visits to the UK.
We are also grateful to Prof. V.Yu. Reshetnyak for facilitating and
encouraging our collaboration.
 \end{acknowledgments}

\appendix

\section{Fluctuation spectrum and stability conditions}
\label{sec:stab-fluct-spectr}

In this appendix we comment on the eigenvalue problem written in a
form similar to Eqs.~\eqref{eq:str-EV}-\eqref{eq:w-BC}
\begin{align}
&
 \label{eq:EV}
  \bigl[\prt{\tau}^2+\lambda\bigr]\,X_{\lambda}(\tau) = 0,
\\
&
\label{eq:BC}
\Bigl[\pm\prt{\tau}
X_{\lambda}+w_{\pm} X_{\lambda}
\Bigr]_{\tau=\pm 1}=0,
\end{align}
where $\lambda$ is the eigenvalue and $X_{\lambda}(\tau)$ is the eigenfunction.
Our task is to derive the conditions which ensure positive
definiteness of the eigenvalues.

To this end we consider the case of negative eigenvalues with
$\lambda=-\kappa^2$ and substitute the general solution of
Eq.~\eqref{eq:EV}
\begin{equation}
  \label{eq:gen-sol}
  X_{\lambda}(\tau) = A_{\lambda}\sinh\kappa\tau + B_{\lambda}\cosh\kappa\tau
\end{equation}
into the boundary conditions~\eqref{eq:BC}. This yields a
homogeneous system of two linear algebraic equations for the
intergration constants $A_{\lambda}$ and $B_{\lambda}$. The system
can be written in matrix form as follows
\begin{equation}
  \label{eq:system}
  \vc{H}\cdot \begin{pmatrix}A_{\lambda}\\ B_{\lambda}\end{pmatrix}=
\begin{pmatrix}0\\ 0\end{pmatrix},
\end{equation}
where
\begin{equation}
  \label{eq:coeff-matrix}
\vc{H}=
  \begin{pmatrix}
\kappa\cosh\kappa + w_{+}\sinh\kappa &
\kappa\sinh\kappa+w_{+}\cosh\kappa\\
-\kappa\cosh\kappa - w_{-}\sinh\kappa &
\kappa\sinh\kappa+w_{-}\cosh\kappa
\end{pmatrix}.
\end{equation}
Non-zero solutions of Eq.~\eqref{eq:system} exist only if the
determinant of the coefficient matrix $\vc{H}$ vanishes,
$\det\vc{H}=0$. For the matrix~\eqref{eq:coeff-matrix}, this
yields a transcendental equation
\begin{equation}
  \label{eq:kappa-eq0}
  \kappa^2\sinh 2\kappa+(w_{+}+w_{-})\kappa\cosh 2\kappa +w_{+}w_{-}\sinh 2\kappa = 0
\end{equation}
whose roots determine the negative eigenvalues through the
relation $\lambda=-\kappa^2$.

Eq.~\eqref{eq:kappa-eq0} can be conveniently recast into the form
\begin{equation}
  \label{eq:x-eq}
  f(x)\equiv x^2+a x\coth x = b,
\end{equation}
where $x\equiv 2\kappa$, $a\equiv 2(w_{+}+w_{-})$ and
$b\equiv -4w_{+}w_{-}$. It is now not difficult to see that the
inequality
\begin{equation}
  \label{eq:stab1}
  b<\min_{x\ge 0} f(x)\equiv \min_{x\ge 0}(x^2+a x\coth x)
\end{equation}
provides the condition for the eigenvalues to be positive.

Eq.~\eqref{eq:stab1} combined with the relation $b>-a^2/4$
can be analyzed using elementary methods. The final result
\begin{equation}
  \label{eq:stab-final}
  b<a,\quad a>0
\end{equation}
immediately leads to the inequalities
\begin{equation}
  \label{eq:stab-cond}
  w_{+}+w_{-}+2w_{+}w_{-}>0,\quad
w_{+}+w_{-}>0,
\end{equation}
which ensure positive definiteness of the spectrum. In the
semi-strong limit, $w_{-}\to\infty$, the
conditions~\eqref{eq:stab-cond} assume the following simplified
form:
\begin{equation}
  \label{eq:semi-appen}
  1+2w_{+}>0.
\end{equation}

Our final remark is that changing $\kappa$ to $i\kappa$
in Eq.~\eqref{eq:kappa-eq0} gives the equation
\begin{equation}
  \label{eq:kappa-eq}
  (w_{+}w_{-}-\kappa^2)\sin 2\kappa+(w_{+}+w_{-})\kappa\cos 2\kappa = 0.
\end{equation}
The roots of this equation determine positive eigenvalues $\lambda=\kappa^2$.
In the strong anchoring
limit, $w_{\pm}\to\infty$, Eq.~\eqref{eq:kappa-eq} takes the
well known form $\sin 2\kappa = 0$ leading to the relation~\eqref{eq:lmb-m-str}.


\end{document}